\documentclass[10pt,sigconf]{acmart}
\usepackage[utf8]{inputenc}
\usepackage{graphicx,url}
\usepackage{xspace}
\usepackage{enumitem}
\usepackage{tikz}
\usepackage{multirow}
\usepackage{multicol}
\usepackage{amsmath}
\DeclareMathOperator*{\argmin}{argmin}

\usetikzlibrary{arrows.meta}
\usepackage{balance}

\newcommand{\stitle}[1]{\vspace{1ex}\noindent{\bf #1}}
\newcommand{\ititle}[1]{\vspace{0.8ex}\noindent{{\em #1}}}
\newcommand{\rev}[1]{\textcolor{black}{{#1}}}
\newcommand{\final}[1]{{{#1}}}
\newcommand{\system}{{\sc EmbDI}\xspace} 
\usepackage{algorithm}
\usepackage[noend]{algorithmic}
\usepackage{color, colortbl}

\begin{document}
\title{
    Creating Embeddings of Heterogeneous Relational Datasets for Data Integration Tasks
}

\sloppy

\copyrightyear{2020}
\acmYear{2020}
\setcopyright{acmcopyright}\acmConference[SIGMOD'20]{Proceedings of the 2020
ACM SIGMOD International Conference on Management of Data}{June 14--19,
2020}{Portland, OR, USA}
\acmBooktitle{Proceedings of the 2020 ACM SIGMOD International Conference on
Management of Data (SIGMOD'20), June 14--19, 2020, Portland, OR, USA}
\acmPrice{15.00}
\acmDOI{10.1145/3318464.3389742}
\acmISBN{978-1-4503-6735-6/20/06}



\author{Riccardo Cappuzzo}
\email{cappuzzo@eurecom.fr}
\affiliation{%
  \institution{EURECOM}
}

\author{Paolo Papotti}
\email{papotti@eurecom.fr}
\affiliation{%
  \institution{EURECOM}
}

\author{Saravanan Thirumuruganathan}
\email{sthirumuruganathan@hbku.edu.qa}
\affiliation{%
  \institution{QCRI, HBKU}
}

 \fancyhead{}

\begin{abstract}
Deep learning based techniques have been recently used with promising results for data integration problems.
Some methods directly use \textit{pre-trained} embeddings that were trained on a large corpus such as Wikipedia.
However, they may not always be an appropriate choice for enterprise datasets with custom vocabulary.
Other methods adapt techniques from
natural language processing to obtain embeddings for the enterprise's relational data.
However, this approach blindly treats a tuple as a sentence, thus losing a large amount of contextual information present in the tuple.

We propose algorithms for obtaining \emph{local embeddings} that are effective for data integration tasks on relational databases. We make four major contributions. First, we describe a compact graph-based representation that allows the specification of a rich set of relationships inherent in the  relational world. Second, we propose how to derive sentences from such a graph that effectively ``describe" the similarity across elements (tokens, attributes, rows) in the two datasets. The embeddings are learned based on such sentences. Third, we propose effective optimization to improve the quality of the learned embeddings and the performance of integration tasks.
Finally, we propose a diverse collection of criteria to evaluate relational embeddings and perform an extensive set of experiments validating them against multiple baseline methods.
Our experiments show that our framework, \system, produces meaningful results for data integration tasks
such as schema matching and entity resolution both in supervised and
 unsupervised settings. 
\end{abstract}


\maketitle

\section{Introduction}
\label{sec:introduction}

Data in an enterprise is often scattered across information silos.
The problem of data integration concerns the combination of information from
heterogeneous relational data sources~\cite{GolshanHMT17}.
It is a challenging first step before data analytics can be performed to extract value from data.
Unfortunately, it is also an expensive task for humans~\cite{rattenbury2017principles}.
An often cited statistic is that data scientists spend 80\% of their time
integrating and curating their data~\cite{crowdSurvey}.
Due to its importance, the problem of data integration has been studied extensively by the database community.
Traditional approaches require substantial effort from domain scientists
to generate features and labeled data or domain specific rules~\cite{GolshanHMT17}.
There has been increasing interest in achieving accurate data integration with dramatically less human effort.

\subsection{Word Embeddings for Data Integration}
Embeddings have been successfully used for data integration tasks such as
entity resolution~\cite{deeper,deepmatcher,ZhaoH19,CakalMA19,kasai2019low,thirumuruganathan2018reuse},
schema matching~\cite{seepingSemantics,miller2018making,koutrasrema},
identification of related concepts~\cite{termite}, and data curation in general~\cite{Hulsebos:2019,dcVision}.
Typically, these works fall into two dominant paradigms based on
how they obtain word embeddings. The first is to 
reuse \textit{pre-trained} word embeddings
computed for a given task. The second is to 
build \textit{local} word embeddings 
that are specific to the dataset.
These methods treat each tuple as a sentence by
reusing the same techniques for learning word embeddings employed in natural language processing.

\begin{figure*}[h]
\hspace{-2ex}
\includegraphics[scale=.75]{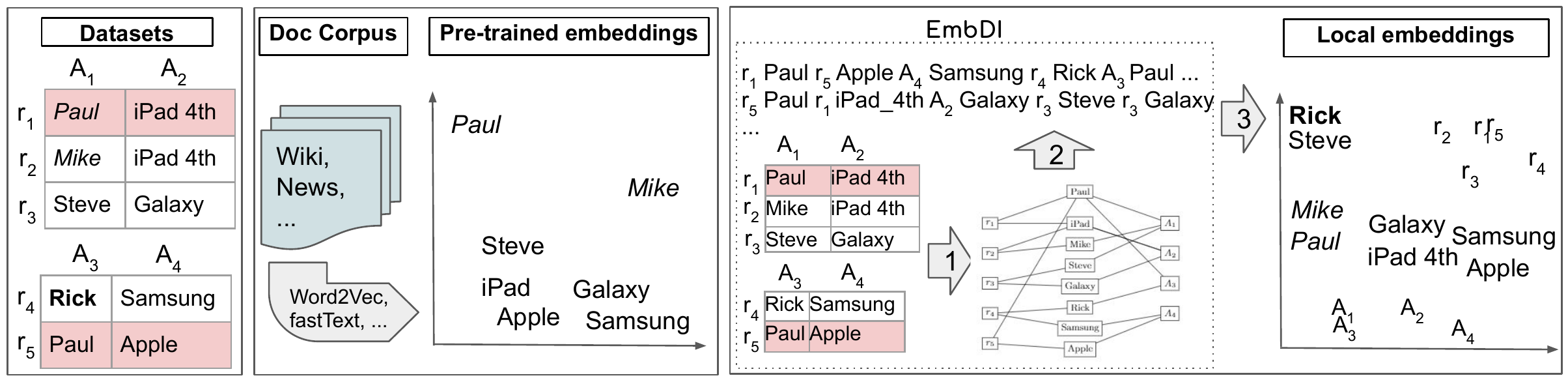}
\caption{Illustration of a simplified vector space learned from text (prior approaches) and from data (\system).}
\label{fig:universe}
\end{figure*}

However, both approaches fall short in some circumstances.
Enterprise datasets tend to contain custom vocabulary.
For example, consider the small datasets reported in the left-hand side of Figure~\ref{fig:universe}. The pre-trained embeddings do not capture the semantics expressed by these datasets and do not contain embeddings for the word ``Rick''.
Approaches that treat a tuple as a sentence miss a number of signals such as attribute boundaries, integrity constraints, and so on. Moreover, existing approaches do not consider the generation of embeddings from heterogeneous datasets, with different attributes and alternative value formats. These observations motivate the generation of \textit{local} embeddings for the \textit{relational} datasets at hand.

\subsection{Local Embeddings for Data Integration}
We advocate for the design of local embeddings that leverage both
the relational nature of the data and the downstream task of data integration.

\noindent \textit{Tuples are not sentences.} Simply adapting embedding techniques originally developed for textual data ignores the richer set of semantics inherent in \textit{relational} data.
Consider a cell value $t[A_i]$ of an attribute $A_i$ in tuple $t$, e.g., ``Mike'' in the first relation from the top.
Conceptually, it has a semantic connections with both other attributes of tuple $t$
(such as ``iPad 4th'')
and other values from the domain of attribute $A_i$ (such as ``Paul'').
Existing embedding techniques cannot such semantic connections.

\noindent \textit{Embedding generation must span different datasets.}
Embeddings must be trained using heterogeneous datasets, so that
they can meaningfully leverage and surface similarity across 
data sources.
A notion of similarity
between different types of entities, such as tuples and attributes, must be developed. 
Tuple-tuple and attribute-attribute similarity are important features for entity resolution and schema matching.

There are multiple challenges to overcome. First, it is not clear how to encode the semantics of the relational datasets into the embedding learning process. Second, datasets may share very limited amount of information, have radically different schemas, and contain a different number of tuples. Finally, datasets are often incomplete and noisy. The learning process is affected by low information quality, generating embeddings that do not correctly represent the semantics of the data.




\subsection{Contributions}
We introduce \system, a framework for building relational, local embeddings for data integration that introduces a number of innovations to overcome the challenges above.
We identify crucial components and propose effective algorithms for instantiating each of them.
\system is designed to be modular so that any one can customize it by plugging in other algorithms
and benefit from the continuing improvements from the deep learning and the database communities.
The right-hand side of Figure~\ref{fig:universe} shows the main steps in our solution.

\stitle{1. Graph Construction.}
We leverage a compact tripartite graph-based representation of relational datasets that can effectively represent a rich set of syntactic and semantic relationships between cell values.
Specifically, we use a heterogeneous graph with three types of nodes.
\emph{Token} nodes correspond to the unique values found in the dataset.
\emph{Record Id} nodes (RIDs) represent a unique token for each tuple. 
\emph{Column Id} nodes (CIDs) represent a unique token for each column/attribute.
These nodes are connected by edges based on the structural relationships in the schema.
This graph is a compact representation of the original datasets that highlights overlap and explicitly represent the primitives for data integration tasks, i.e., records and attributes.

\stitle{2. Embedding Construction.}
We formulate the problem of obtaining local embeddings for relational data as a graph embeddings generation problem.
We use random walks to quantify the similarity between neighboring nodes and to exploit metadata such as tuple and attribute IDs. This method ensures that nodes that share similar neighborhoods will be in close proximity in the final embeddings space.
The corpus that is used to train our local embeddings is generated by materializing these random walks.



\stitle{3. Optimizations.}
Learning embeddings can be a difficult task in the presence of noisy and incomplete heterogeneous datasets. For this reason, we introduce an array of optimization techniques that handle difficult cases and enable refinement of the generated embeddings.
The flexibility of the graph enables us to naturally represent external information, such as data dictionaries, to merge values in different formats, and data dependencies, to impute values and identify errors.
We propose optimizations to handle imbalance in the datasets' size and the presence of numerical values (usually ignored in textual word embeddings).

\stitle{Experimental Results.}
We propose an extensive set of desiderata for evaluating relational embeddings for data integration.
Specifically, our evaluation focuses on three major dimensions that measure how well do the embeddings
(a) learn the tuple-, attribute- and constraint-based relationships in the data,
(b) learn integration specific information such as tuple-tuple and attribute-attribute similarities,
and
(c) improve the behavior of DL-based data integration algorithms.
As we shall show in the experiments, our proposed algorithms perform well on each of these dimensions.

\stitle{Outline.}
Section~\ref{sec:background} introduces background about embeddings and data integration.
Section~\ref{sec:motivatingExample} shows a motivating example that highlights the limitations of prior approaches and identifies a set of desiderata for relational embeddings.
Section~\ref{sec:overview} details the major components of the framework.
Section~\ref{sec:extensions} presents our optimizations to handle data imbalance, missing values, and external information.
Section~\ref{sec:embeddingsForIntegration} describes how we use embeddings for data integration tasks.
Section~\ref{sec:experiments} 
reports extensive experiments validating our approach. We conclude in Section~\ref{sec:conclusions} with some promising next steps.

\section{Background}
\label{sec:background}

\stitle{Embeddings.}
Embeddings 
map an entity such as a word to a high dimensional real valued vector. The mapping is performed in such a way that the geometric relation between the vectors of two entities represents the co-occurrence/semantic relationship between them. 
%
Algorithms used to learn embeddings 
rely on the notion of ``neighborhood'': intuitively, if two entities are similar, they frequently belong to the same contextually defined neighborhood. When this occurs, the embeddings generation algorithm will try to force the vectors that represent these two entities to be close to each other in the resulting vector space.

\stitle{Word Embeddings}~\cite{turian2010word,BojanowskiGJM16}
are trained on a large corpus of text and
produce as output a vector space where each word in the corpus is represented by a real valued vector.
Usually, the generated vector space has either 100 or 300 dimensions.
The vectors for words that occur in similar context -- such as SIGMOD and VLDB -- are in close proximity to each other.
Popular architectures for learning embeddings include continuous bag-of-words (CBOW) or skip-gram (SG).
Recent approaches rely on using the context of word to obtain a contextual word embedding~\cite{devlin2018bert,abs-1802-05365}.

\stitle{Node Embeddings.}
Intuitively, node embeddings~\cite{node2vec} map nodes to a high dimensional vector space so that the likelihood of preserving node neighborhoods is maximized. One way to achieve this is by performing random walks starting from each node to define an appropriate neighborhood. Popular node embeddings are often based on the skip-gram model, since it maximizes the probability of observing a node's neighborhood given its embedding. By varying the type of random walks used, one can obtain diverse types of embeddings~\cite{ChenPHS17}.

\stitle{Embeddings for Relational Datasets.}
The pioneering work of~\cite{ibmRelEmb} was the first to apply embedding techniques
for extracting latent information from relation data.
Recent extensions~\cite{bordawekar2019exploiting,bordawekar2017cognitive} leverage the learned embeddings
to develop a ``cognitive'' database system with sophisticated functionality for
answering complex  semantic, reasoning and predictive queries.
Termite~\cite{termite} seeks to project tokens from structured and unstructured data into a common representational space that could then be used for identifying related concepts through its Termite-Join approach.
Freddy~\cite{freddy} and RetroLive~\cite{nikulskiretrolive} produce relational embeddings
that combine relational and semantic information through a retrofitting strategy.
There has been prior work that learn embeddings for specific tasks like
entity matching (such as DeepER~\cite{deeper} and DeepMatcher~\cite{deepmatcher}) and
schema matching (Rema~\cite{koutrasrema}).
Our goal is to learn relational embeddings that is tailored for data integration
and can be used for multiple tasks.


\section{Motivating Example}
\label{sec:motivatingExample}
In this section, we discuss an illustrative example that highlights the weaknesses of current approaches and motivates us to design a new approach for relational, local embedding.

Consider the scenario where one utilizes popular pre-trained embeddings 
such as word2vec, GloVe, or fastText. 
Figure~\ref{fig:universe} shows a hypothetical filtered vector spaces for the tokens in an example with two small customer datasets.
We observe that the pre-trained embeddings suffer from a number of issues when we use them to model the two relations.
\begin{enumerate}[leftmargin=*]
    \item A number of words, such as ``Rick'', in the dataset are not in the pre-trained embedding.
    This is especially problematic for enterprise datasets where tokens are often unique and not found in pre-trained embeddings.
    \item Embeddings might contain geometric relationships that exist in the corpus they were trained on, but that are missing in the relational data.
    For example, the embedding for token ``Steve'' is closer to tokens ``iPad'' and ``Apple'' even though it is not implied in the data.
    \item Relationships that do occur in the data, such as between tokens ``Paul'' and ``Mike'',  are not observed in the pre-trained vector space.
\end{enumerate}

Naturally, learning local embeddings from the relational data often produces better results. However, computing embeddings for non integrated data sources is a non trivial task.
This becomes especially challenging in settings where data is scattered over different datasets with heterogeneous structures, different formats, and only partially overlapping content.
Prior approaches express such datasets as sentences that can be consumed by existing word embedding methods.
However, we find that these solutions are still sub-optimal for downstream data integration tasks.

\stitle{Technical Challenges.}
We enumerate four challenges that must be overcome to obtain effective embeddings.

\ititle{1. Incorporating Relational Semantics.}
Relational data exhibits a rich set of semantics.
Relational data also follows set semantics where there is no natural ordering of attributes.
Representing the tuple as a single sentence is simplistic and often not expressive enough for these signals.

\ititle{2. Handling Lack of Redundancy.}
A key reason for the success of word embeddings is that they are trained on large corpora where there are adequate redundancies and co-occurrence to learn relationships.
However, databases are often normalized to remove redundant information.
This has an especially deleterious impact on the quality of learned embeddings. 
Rare words, which are very common in relational data, are typically ignored by word embedding methods. 

\ititle{3. Handling Multiple Datasets.}
We cannot assume that each of the datasets have the same set of attributes,
or that there is sufficient overlapping values in the tuples,
or even that there is a common dictionary for the same attribute.

\ititle{4. Handling Hierarchical Data.}
Databases are inherently hierarchical, with entities such as cell values, tuples, attributes, dataset and so on.
Incorporating these hierarchical units as first class citizens in embedding training is a major challenge.






\section{Constructing Local Relational Embeddings}
\label{sec:overview}
In this section, we provide a description of our approach and how these design choices address the aforementioned technical challenges.
Our framework, \system, consists of three major components, as depicted in the right-hand side of Figure~\ref{fig:universe}.

\begin{enumerate}[leftmargin=*]
    \item In the \emph{Graph Construction} stage,
        we process the relational dataset and transform it to a compact tripartite graph that
        encodes various relationships inherent in it.
        Tuple and attribute ids are treated as first class citizens.
    \item Given this graph, the next step is \emph{Sentence Construction}
        through the use of biased random walks.
        These walks are carefully constructed to avoid common issues such as rare words and imbalance in vocabulary sizes. This produces as output a series of sentences.
    \item In \emph{Embedding Construction},
        the corpus of sentences is passed to an algorithm for learning word embeddings.
        Depending on available external information, 
        we perform optimizations to the graph and the workflow to improve the embeddings' quality.
\end{enumerate}

\subsection{Graph Construction}
\label{subsec:graphConstruction}

\stitle{Why construct a Graph?}
Prior approaches for local embeddings seek to directly apply an existing
word embedding algorithm on the relational dataset.
Intuitively, all tuples in a relation are modeled as sentences by breaking the attribute boundaries.
The collection of sentences for each tuple in the relation then makes up the corpus,
which is then used to train the embedding.
This approach produces embeddings that are customized to that dataset, but
it also ignores signals that are inherent in relational data.
We represent the relational data as a graph, thus enabling a more expressive representation with a number of advantages.
First, it elegantly handles many of the various relationships between entities that are common in relational datasets.
Second, it provides a straightforward way to incorporate external information
such as ``two tokens are synonyms of each other''.
Finally, when multiple relations are involved, a graph representation enables a unified view over the different datasets that is  invaluable for learning embeddings for data integration.

\stitle{Simple Approaches.}
Consider a relation $R$ with attributes $\{A_1, A_2, \ldots, A_m\}$.
Let $t$ be an arbitrary tuple and $t[A_i]$ the value of attribute $A_i$ for tuple $t$.
A naive approach is to create a chain graph where tokens corresponding to adjacent attributes such as $t[A_i]$ and $t[A_{i+1}]$ are connected.
This will result in $m$ edges for each tuple.
Of course, if two different tuples share the same token, then they will reuse the same node.
However, relational algebra is based on set semantics, where the attributes do not have an inherent order.
So, simplistically connecting adjacent attributes is doomed to fail.
Another extreme is to create a complete subgraph, where an edge exists between all possible pairs of $t[A_i]$ and $t[A_{i+1}]$.
Clearly, this will result in ${m \choose 2}$ edges per tuple.
This approach results in the number of edges is quadratic in the number of attributes and ignores other token relationships
such as ``token $t_1$ and token $t_2$ belong to the same attribute''.

\stitle{Relational Data as Heterogeneous Graph.}
We propose a heterogeneous graph with three types of nodes.
\rev{\emph{Token} nodes correspond to information found in the dataset (i.e. the content of each cell in the relation). Multi-word tokens may be represented as a single entity, get split over multiple nodes or use a mix of the two strategies. We describe the effect of each strategy more in depth in Section \ref{sec:experiments}.} 
\emph{Record Id} nodes (RIDs) represent each tuple in the dataset,
\emph{Column Id} nodes (CIDs) represent each column/attribute.
These nodes are connected by 
edges according to the structural relationships in the schema.
\rev{This representation can produce a vector for all RIDs (CIDs) rather than representing them by combining the vectors of the values in each tuple (column).
}

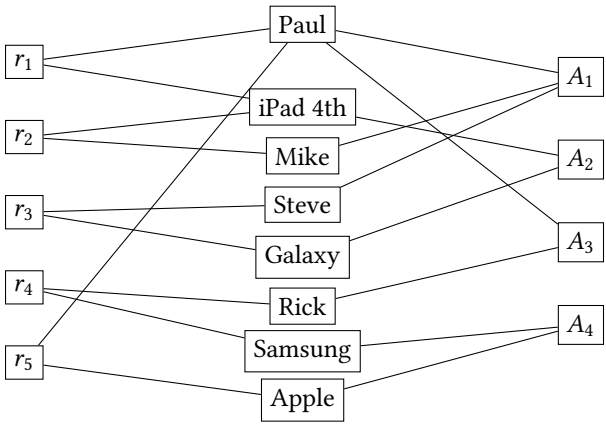
\begin{figure}[th]
\centering
\begin{tikzpicture}
\begin{scope}[every node/.style={draw}]
    \node (r1) at (-1.2,2.5) {$r_1$};
    \node (r2) at (-1.2,1.5) {$r_2$};
    \node (r3) at (-1.2,0.5) {$r_3$};
    \node (r4) at (-1.2,-0.5) {$r_4$};
    \node (r5) at (-1.2,-1.5) {$r_5$};
    \node (paul) at (2.5,3.0) {Paul};
    \node (4th) at (2.5,1.9) {iPad 4th};
    \node (mike) at (2.5,1.25) {Mike};
    \node (steve) at (2.5,0.6) {Steve};
    \node (galaxy) at (2.5,-0.1) {Galaxy};
    \node (rick) at (2.5,-0.75) {Rick};
    \node (samsung) at (2.5,-1.35) {Samsung};
    \node (apple) at (2.5,-2) {Apple};
    \node (A1) at (6.2,2.3) {$A_1$};
    \node (A2) at (6.2,1.2) {$A_2$};
    \node (A3) at (6.2,0.1) {$A_3$};
    \node (A4) at (6.2,-1) {$A_4$};
\end{scope}

\begin{scope}[>={Stealth[black]},
              every edge/.style={draw=black}]
    \path [-] (r1) edge (paul);
        \path [-] (r5) edge (paul);
    \path [-] (r1) edge (4th);
    \path [-] (r2) edge (mike);
    \path [-] (r2) edge (4th);
    \path [-] (r3) edge (steve);
    \path [-] (r3) edge (galaxy);
    \path [-] (r4) edge (rick);
    \path [-] (r4) edge (samsung);
    \path [-] (r5) edge (apple);
    \path [-] (A1) edge (paul);
    \path [-] (A1) edge (mike);
    \path [-] (A2) edge (4th);
    \path [-] (A1) edge (steve);
    \path [-] (A2) edge (galaxy);
    \path [-] (A3) edge (rick);
    \path [-] (A4) edge (samsung);
    \path [-] (A3) edge (paul);
    \path [-] (A4) edge (apple);
\end{scope}
\end{tikzpicture}
\caption{The graph for the two tables in Figure~\ref{fig:universe}.}
\label{fig:graph_example}
\end{figure}

Consider a tuple $t$ with RID $r_t$.
Then, nodes for tokens corresponding to $t[A_1], \ldots, t[A_m]$ are connected to the node $r_t$.
Similarly, all the tokens belonging to a specific attribute $A_i$ are connected to the corresponding CID, say $c_i$.
This construction is generic enough to be augmented with other types of relationships.
Also, if we know that two tokens are synonyms (e.g. via wordnet),
this information could be incorporated by reusing the same node for both tokens.
Note that a token could belong to different record ids and column ids
when two different tuples/attributes share the same token.
Numerical values are rounded to a number of significant figures decided by the user, then they are assigned a node like regular categorical values; null values are not represented in the graph.
We discuss more sophisticated approaches for handling numeric, noisy, and null values in Section~\ref{sec:extensions}.

\rev{Algorithm \ref{alg:graphGeneration} shows the operations performed during the graph creation with hybrid representation of multi-word tokens}.
Figure~\ref{fig:graph_example} shows a graph constructed for the datasets in Figure~\ref{fig:universe}.
Note that this could be considered as a variant of tripartite graph.
A key advantage of this choice is that it has the same expressive power as the complete sub-graph approach, while requiring orders of magnitude fewer edges.

\begin{algorithm}[ht]
        \caption{\rev{GenerateTripartiteGraph}}
        \label{alg:graphGeneration}
        \begin{algorithmic}[]
        \STATE \textbf{Input}: relational dataset $D$ 
        \STATE let $G$ = empty graph
        \FORALL{$c_i$ in columns($D$)}
            \STATE G.addNode($c_i$)
        \ENDFOR
        \FORALL{$r_i$ in rows($D$)}
            \STATE G.addNode($R_i$) \quad //$R_i$ is the record id of $r_i$ 
            \FORALL{value $v_k$ in $r_i$}
                \IF{$v_k$ is multi-word}
                    \FORALL{word in tokenize($v_k$)}
                        \STATE G.addNode(word)
                        \STATE G.addEdge(word, $R_i$), G.addEdge(word, $c_k$)
                    \ENDFOR
                \ELSIF{$v_k$ is single-word}
                    \STATE G.addNode($v_k$)
                    \STATE G.addEdge($v_k$, $R_i$), G.addEdge($v_k$, $c_k$)
                \ENDIF
            \ENDFOR
        \ENDFOR
        \STATE \textbf{Output}:  graph $G$
    \end{algorithmic}
\end{algorithm}

\subsection{Sentence Construction}
\label{subsec:randomWalks}

\stitle{Graph Traversal by Random Walks.}
To generate the distributed representation of each node in the graph, we produce a large number of random walks and gather them in a training corpus where each random walk will correspond to a sentence.
Using graphs and random walks allows us to have a richer and more diverse set of neighborhoods than what would be possible by encoding a tuple as a \emph{single} sentence.
For example, a walk starting from node `Paul' could go to node $A_3$, and then to node `Rick'.
This walk implicitly defines the neighborhood based on attribute co-occurrence.
Similarly, the walk from `Paul' could have gone to `$r_5$' and then to `Apple', incorporating the row level relationships.
Our approach is agnostic to the specific type of random walk used, with different choices yielding different embeddings.
For example, one could design random walks that are biased towards other nodes belonging to the same tuple, or towards rare nodes.
\rev{To better represent all nodes, we assign a ``budget'' of random walks to each of them and guarantee that all nodes will be the starting point of at least as many random walks as their budget.
After choosing the starting point $T_i$, the random walk is generated by choosing a neighboring RID of $T_i$, $R_j$. The next step in the random walk will then be chosen at random among all neighbors of node $R_j$, for example by moving on $C_a$. Then, a new neighbor of $C_a$ will be chosen and the process will continue until the random walk has reached the target length.
}
We use uniform random walks in most of our experiments to guarantee good execution times on large datasets, while providing high quality results. We compare alternative random walks in the experiments.

\begin{algorithm}
\begin{algorithmic}[]
        \caption{\rev{GenerateRandomWalk}}
        \label{alg:rwGeneration}
        \STATE \textbf{Input}: starting node $n_j$, random walk length $l$
        \STATE $r_j$ = findNeighboringRID($n_j$)
        \STATE $W$ = seq($r_j$, $n_j$)
        \STATE currentNode = $n_j$
        \WHILE{length($W$) < $l$}
            \STATE	nextNode = findRandomNeighbor(currentNode)
            \STATE  $W$.add(nextNode)
            \STATE currentNode = nextNode
    	\ENDWHILE
    \STATE \textbf{Output}: walk $W$
\end{algorithmic}
\end{algorithm}


\stitle{From Walks to Sentences.}
It is important to note that the path on the graph represented by the random walk does not necessarily reflect the sentence that will be inserted in the training corpus.
For example, a possible random walk could be the following: $R_a T_b R_c T_d C_e T_f C_g T_h$, where $T_{*}, R_{*}, C_{*}$ correspond to nodes of type tokens, record ids, and column ids, respectively.
{We note that the random walks include nodes corresponding to RIDs and CIDs. We noticed that the presence  (or absence)  of CIDs and RIDs in the sentences that build the training corpus has large effects on the data integration performance of the algorithm. Indeed, we observe that treating these as first order citizens, we can represent them as points in the vector space in the same way as any other token.}
For example, two nodes corresponding to different attributes might co-occur in many random walks, resulting in embeddings that are closer to each other: this may imply that these two attributes represent similar information.
A similar phenomenon could also be obtained for tuple embeddings.
A number of prior approaches such as DeepER~\cite{deeper} or DeepMatcher~\cite{deepmatcher}
only learn embeddings for tokens and then obtain embeddings for tuples
by averaging them or combining by using a RNN.
%
The use of our random walks as sentences provides additional information about the neighborhood of each node, which would not be so easily obtained by using only the structured data format.

\subsection{Embedding Construction}
\label{subsec:embeddingConstruction}
The generated sentences are then pooled together to build a corpus that is used to train the embeddings algorithm.
Our approach is agnostic to the actual word embedding algorithm used.
We piggyback on the plethora of effective embeddings algorithms such as word2vec, GloVe, fastText, and so on. 
Every year, improved embedding training algorithms are released, and this has a transitive effect on our approach.
Broadly, these techniques can be categorized as word-based (such as word2vec) or character-based (such as fastText).
%
We discuss the hyperparameters for embedding algorithms such as learning method (either CBOW or Skip-Gram), dimensionality of the embeddings, and size of context window in Section~\ref{sec:experiments}.


\subsection{Algorithm So Far}
\rev{
Algorithm~\ref{alg:basicAlgo} provides the pseudocode for learning the local and relational embeddings based on our discussion.
In the next section, we discuss a number of practical improvements to this basic algorithm.}

\begin{algorithm}[h]
        \caption{\rev{Meta Algorithm for \system}}
        \label{alg:basicAlgo}
        \begin{algorithmic}[1]
                \STATE \textbf{Input:} relational datasets $D$, number of random walks $n_{walks}$, number of nodes $n_{nodes}$
                    \STATE $W$ = [] 
                    \STATE $G$ = GenerateTripartiteGraph($D$)
                    \FORALL{ $n_j$ $\in$ $nodes(G)$}
                        \FOR{i = 1 to ($n_{walks}/n_{nodes}$)}
                            \STATE $w_i$ = GenerateRandomWalk($n_j$)
                            \STATE $W$.add($w_i$)
                        \ENDFOR
                    \ENDFOR
                    \STATE $E$ = GenerateEmbeddings($W$)
                \STATE \textbf{Output:} 
                 Local relational embeddings $E$
        \end{algorithmic}
\end{algorithm}


\section{Improving Local Embeddings}
\label{sec:extensions}
In this section, we discuss a number of challenging issues that occur when applying \system in practice.

\subsection{Handling Imbalanced Relations}
In a real-world scenario, there often are multiple relations and local embeddings must be learned for each of them.
For a single relation, one can simply perform multiple random walks from each token node.
This approach directly ameliorates the issue of infrequent words that plagues word embedding approaches, by guaranteeing that even rare words will appear frequently enough to be properly represented. 
A further complication arises when one relation 
contains
many more nodes than the other.
If we perform an equal amount of random walks starting from each node,
the signals from the larger dataset might overwhelm those coming from the smaller dataset.
We found that an effective heuristic is to
start random walks only from nodes that co-occur in both datasets.
This approach often produces sentences where the proportion of larger and smaller datasets is comparable.
Furthermore, these nodes also happen to be the most informative ones as they connect two relationships and often quite useful for integrating these two relations.
Even with datasets with a minimum amount of overlap (less than 2\%), this approach ensures adequate coverage of all nodes and minimizes the issues due to relation imbalance.

The overlapping tokens are the bridge between the two datasets to be integrated. To maximize their impact in the embedding creation, one could start every sentence with a RID or CID, randomly picked from those connected to the token at hand. This small change in the random walk creation affects the results by creating evidence of similarity for the corresponding rows and columns.

\ititle{Example.} Assume that node token $T_a$ appears in two rows $R_a$ and $R_b$ over two large datasets. Since the token is rare,  it will appear most likely only once as the first node in the walk, therefore the embedding algorithm will only see it in few patterns, such as $T_a R_b T_c$ or $T_a C_d T_e$. To improve the modeling of the $T_a$ we start the sentence with a RID or CID connected to $T_a$, such as $C_d T_a C_c$ and $R_a T_a R_b$. This way, even if the token is rare, it gives strong signals that the attributes and the row that contain it are related.

\subsection{Handling Missing and Noisy Data}
\label{sec:skolem}
Many real-world datasets contain a large amount of missing data, so any effective approach for learning embeddings must have a cogent strategy for this scenario.
The ideal approach employs imputation techniques to minimize the number of missing values.
Unfortunately, this might not always be possible, since algorithms for imputation and data repair often do not provide good results in a relational setting.
Prior approaches for learning relational embeddings skip missing values when computing embeddings.
However, this approach is often counter-productive as missing data can be an indication of systemic error.
Approaches where all missing values are treated as if they were the same entity (so one node for all nulls), or unique entities (individual nodes for each null) are not appropriate.
\rev{The first approach creates a super node to store all NULL values, which has multiple negative effects on the result and produces no benefit.}
The second approach creates a unique node for each NULL: this does not cause any issues, but  does not provide any additional information either. Moreover, if the number of NULLs is large, this approach increases the processing time without any commensurate benefit.

We propose a simple mechanism to use classical database techniques such as Skolemization~\cite{HullY90} to handle missing data.
Approaches for data repairs~\cite{XuBook19} are very accurate in identifying the errors,
but struggle to identify the correct updated value~\cite{AbedjanCDFIOPST16,ArocenaGMMPS15}.
When there is no certain update to make, most methods put a \textit{placeholder},
like a variable or the output of a function that is related to Skolemization.
Our model is able to naturally consume and model these placeholders to obtain better embeddings.
Hence, the data repairing task could be used to address both missing and noisy values.

Consider the scenario with two relations $R_1$ and $R_2$.
Without loss of generality, let us assume that they both have attributes $A_1, A_2, A_3, A_4$.
Suppose there are two tuples:
\begin{equation*}
    R1(a, N_1 , c, N_2) \text{ and } R2(a, b, c', N_3)
\end{equation*}

Here $N_1, N_2, N_3$ denote the null values.
If $A_1$ is the key attribute, we can derive three important updates in the data,
including the creation of two placeholders,
and rewrite the two tuples are follows:
\begin{equation*}
    R1(a, b , X_1, X_2) \text{ and } R2(a, b, X_1, X_2)
\end{equation*}

where $X_1$ models the conflict between $c$ and $c'$ and $X_2$ merges the two nulls.
This reduces the heterogeneity of the data and improves the quality of the embeddings.
Consider also that all occurrences of $c$ and $c'$ are merged in the graph,
even in tuples that do not satisfy the pattern of this functional dependency. 
A single placeholder may end up merging a large number of token
occurrences in the original dataset.

\subsection{Incorporating External Information} \label{sec:external}
\stitle{Node Merging.}
Our graph representation allows one to incorporate external information such as wordnet or other domain specific dictionaries in a seamless manner.
This is an \emph{optional} step to improve the quality of embeddings.
For example, consider two attributes from different relations -- one stores country codes while the other contains complete country names.
If some mapping between these two exists, then we can \emph{merge} the nodes corresponding to, say, Netherlands and NL. The same reasoning applies to tuples (attributes): if trustable information about possible token matches is available, we merge different RIDs (CIDs) in the same node.
Merging of nodes could be achieved by using external functions, such as matchers based on syntactic similarity, pre-trained embeddings, or clustering.
This often increases the number of overlapping tokens across datasets and produces better embeddings for data integration.

\stitle{Node Replacement in Random Walks.}
Merging of nodes is only viable if we are confident that the two tokens refer to the same underlying entity.
In practice, the mapping between two entities is imperfect.
For example, one could have a machine learning algorithm that says that tokens $T_i$ and $T_j$ are similar with confidence of $0.8$.
The extreme approaches of merging the two nodes (such as by applying a fixed threshold)
or ignoring this strong information are both sub-optimal.
We propose the use of a \emph{replacement} strategy where, during the construction of the sentence corpus, token $T_i$ is replaced by $T_j$ (and vice versa) with a probability proportionate to their closeness.
Note that this only affects the sentence construction.
The random walk by itself is not affected.
Specifically, if the random walk is at node $T_i$, it might output $T_j$ in the sentence instead of $T_i$.
However, when choosing the next node, it will only pick the neighbors of node $T_i$.

\stitle{Handling Numeric Data.}
Integer and real-valued attributes are very common in relational data.
A straightforward approach is to treat them as strings, so that each distinct value is assigned to a node in the graph.
However, this simplistic approach does not always work well, as
it ignores geometric relationships between numbers such as the Euclidean distance.
One way to use this distance information is to \emph{replace} two numbers
if they are within a threshold distance.
Unfortunately, identifying an effective threshold is quite challenging in general.
Consider two set of tokens $\{1, 2, 3, \ldots,\}$ and $\{1, 1.00001, 1.00002, \ldots, 2\}$.
In the former, we can plausibly replace $1$ with $2$ while it would not be appropriate in the latter scenario.
We apply an effective heuristic that combines node replacement with
data distribution-aware distance between two numbers.
Typically, most numeric attributes can be approximated by a small number of
distributions, such as Gaussian or Zipfian.
As an example, if a particular attribute is Gaussian, we can efficiently estimate its parameters
-- mean and variance.
Then, given a number $i$, we generate a random number $r$ around $i$ in
accordance with the learned parameters.
If the new random number is part of the domain of the attribute, then we replace $i$ with $r$.

\subsection{Embedding Alignment}
\label{subsec:embeddingAlignment}

\begin{algorithm}[h!]
	\caption{\rev{AlignEmbeddings}}
	\label{alg:embAlign}
	\begin{algorithmic}[1]
	    \STATE \textbf{Input}: relations $\mathbb{R}_1$, $\mathbb{R}_2$, $\mathbb{E}$ = \system(concat($\mathbb{R}_1$, $\mathbb{R}_2$))
        \STATE let $U_i$ be the set of unique words in $\mathbb{R}_i$ $\forall i \in {1,2}$
        \STATE let $\mathcal{A} = U_1 \cap U_2$ 
        \STATE $A~=~\mathbb{E}(w_i)~\forall~w_i \in \mathbb{R}_1$
        \STATE $B~=~\mathbb{E}(w_j)~\forall~w_j \in \mathbb{R}_2$
        \STATE $W^*~=~\argmin_{W, \mathcal{A}} (WA-B)$
        \STATE $A'=W^*A$
        \FORALL{$w_i \in \mathbb{R}_1 \cup \mathbb{R}_2$}
            \IF{$w_i \in \mathbb{R}_1 \cap \mathbb{R}_2$}
                \STATE $\mathbb{E}'(w_i)$ = average($A'(w_i)$, $B(w_i)$)
                \ELSIF{$w_i \in \mathbb{R}_1$}
                \STATE $\mathbb{E}'(w_i)$ = $A'(w_i)$ 
                \ELSE
                \STATE $\mathbb{E}'(w_i)$ = $B(w_i)$
            \ENDIF
        \ENDFOR
        \STATE \textbf{Output}: Aligned embeddings $\mathbb{E}'$
        \end{algorithmic}
\end{algorithm}

Typically, embeddings for multiple relations are trained using two extreme approaches --
either by training embeddings one relation at a time or
by pooling all the relations and training a common space.
The individual approach is more scalable, but misses out on patterns that could be inferred by pooling the data.
The pooled approach 
must ensure that signals from larger relations do not overpower those from smaller ones.
We advocate for a novel  embedding alignment approach, adapted from multilingual translation~\cite{muse}.

We begin by training embeddings each relation individually. 
This may cause RID and CID vectors that represent different instances of the same entity to differ from each other when the datasets share a small number of common tokens.
To mitigate this problem, 
we \emph{align} the embeddings of the values contained by the two datasets that were trained in the initial execution 
by pivoting on the new information, basically changing the vector space that represents one dataset to better match the vector space of the other.
This allows us to better materialize relationships between tokens, even if they do not co-occur in a single relation.
Furthermore, this approach ensures that the geometric relationships between tokens within each individual dataset are retained.

{Assume that we have two relations $\mathbb{R}_1$ and $\mathbb{R}_2$ with adequate overlap, and that $A$ and $B$ represent the embeddings of words in  $\mathbb{R}_1$ and $\mathbb{R}_2$, respectively. 
It is possible to formulate an orthogonal Procrustes problem \cite{muse}~ by seeking a translation matrix $W^*~=~\argmin_{W, \mathcal{A}}(WA-B)$, with $\mathcal{A} = U_1 \cap U_2$ being the intersection of unique values (the \textit{anchors}) in common between the two starting relations. 
Applying the translation matrix $W^*$ to $A$ yields a translated matrix $A'$, which minimizes the distance between anchor points.
To employ this technique in the ER and SM tasks, we use matching CIDs and RIDs in the original embeddings as anchors to perform the rotation. We then match again on the rotated embeddings. 
Algorithm \ref{alg:embAlign} describes the embedding alignment. }

\subsection{Handling Multi-Word Tokens}
\label{subsec:multiWordTokens}
\rev{
Multi-word tokens are common in relational dataset (such as ``Adobe Photoshop CS3''). 
There are a number of ways in which multi-word cells could be tokenized.
One simple option is to treat the entire word sequence as a single token.
The other option is to tokenize the word sequence, compute the word embeddings for each of the tokens, 
and then aggregate these token embeddings to get the embedding for the multi-word cell. 
There are two key problems: how to tokenize a multi-world cell and 
how to aggregate the token embeddings to get the cell embeddings.
There are no simple answers to this problem. 
In some cases, these multi-word tokens contain substrings that would yield additional information if they were represented as stand-alone nodes (in the example above, ``Adobe'' and ``Photoshop'' are likely candidates). Unfortunately, in the general case it is very hard to pinpoint cases where performing the expansion would improve the results; consider a counterexample such as ``Saving Private Ryan'': in this case, we would rather have a single node to represent the movie title as it likely is a ``primary key'' in the dataset and as such would help when performing integration tasks. 
}

\rev{
To mitigate both issues, we found a simple yet effective heuristic that allows us to handle both multi-word tokens and rare tokens at the same time. Instead of representing all unique values in both datasets as nodes, we make a distinction between nodes that are present in both as they already appear, and those that appear only in one dataset. Then, we tokenize the shared tokens and expand those that are not in common. This effectively allows us to extract the information present within multi-word tokens and, possibly, introduce connections that would be missed otherwise. Moreover, representing the common values as unique tokens introduces ``bridges'' between the datasets, which can be exploited during the step of random walks generation to introduce semantic connections that would not be identified otherwise. 
}
\section{Using Embeddings for Integration}
\label{sec:embeddingsForIntegration}
Once the embeddings are trained, they can be used for common data integration tasks.
We now describe \emph{unsupervised} algorithms that employ the  embeddings produced by \system to perform two tasks widely studied in data integration, Schema Matching and Entity Resolution.

\stitle{Schema Matching (SM).}
Traditional approaches rely on grouping attributes based on the value distributions or use other similarity measures.
Recently,~\cite{seepingSemantics} used embeddings to identify relationships between attributes using both syntactic and semantic similarities.
However, they use embeddings only on attribute/relation names and do not consider the instances -- i.e. values taken by the attribute.

\rev{
Algorithm \ref{alg:smAlgo} describes the steps taken to perform schema matching between two attributes by exploiting their cosine distance in the vector space. Consider that, to prevent false positives in the column alignment, we terminate the algorithm after two iterations have been completed, even if some candidate pools may still contain values. }

\begin{algorithm}[h]
        \caption{\rev{Schema Matching} }
        \label{alg:smAlgo}
        \begin{algorithmic}[1]
        \STATE let $\mathcal{C}_1$ be the set of CIDs of dataset $D_1$ and $\mathcal{C}_2$  be the set of CIDs of dataset $D_2$ 
        \STATE let $d(c_i)$ be the list of distances between column $c_i \in \mathcal{C}_1$ and all other columns $c_k \in \mathcal{C}_2$, sorted in ascending order of distance (and viceversa).
        \STATE let $\mathcal{T}$ = $\mathcal{C}_1 \cup \mathcal{C}_1$ be the set of columns to be matched
        \WHILE{$\mathcal{T} \neq \emptyset$}
            \FORALL{$c_k \in \mathcal{T}$}
                \IF{$d(c_k) \neq \emptyset$}
                    \STATE $c'_k$ = findClosest($d(c_k)$)
                    \STATE $c''_k$ = findClosest($d(c'_k)$)
                    \IF{$c''_k$ == $c_k$}
                        \STATE $c_k$ and $c'_k$ are matched
                        \STATE remove $c_k$, $c'_k$ from $\mathcal{T}$
                    \ELSE
                        \STATE removeCandidate($d(c_k)$, $c'_k$)
                        \STATE removeCandidate($d(c'_k)$, $c_k$)
                    \ENDIF
                \ELSE
                \STATE remove $c_k$ from $\mathcal{T}$
                \ENDIF
            \ENDFOR
        \ENDWHILE
        \end{algorithmic}
\end{algorithm}
\label{alg:SM}


\stitle{Entity Resolution (ER).}
Recent works used pre-existing embeddings to represent tuples~\cite{deeper,deepmatcher}.
In contrast, our approach relies on the use of RIDs as nodes in the heterogenous graph. This allows \system to learn better embeddings for the entire record from the data itself, rather than relying on combination methods such as averaging or concatenating the embeddings of the terms in the tuple. 
This information is then used to perform unsupervised ER by computing the distance between RIDs.
We will also discuss in the experiments how one can piggyback on prior supervised approaches by passing the trained embeddings as features to~\cite{deeper,deepmatcher}.

\rev{
Algorithm \ref{alg:erAlgo} describes the steps taken to identify the matches in the Entity Resolution task. We assume that no matches for $R_i$ are present in $D_1$. }

\begin{algorithm}[h]
        \caption{\rev{Entity Resolution} }
        \label{alg:erAlgo}
        \begin{algorithmic}[1]
        \STATE let $\mathcal{R}_1$ be the set of RIDs $\in D_1$
        \STATE let $\mathcal{R}_2$ be the set of RIDs $\in D_2$
        \STATE let $d(r_i)$ be the list of distances between RID $r_i \in \mathcal{R}_i$ and the closest $n_{top}$ RIDs $\in D_j$, with $i \neq j$. 
        \FORALL{$r_i \in D_1 \cup D_2$ }
            \STATE $d(r_i)$ = findClosest($r_i$, $n_{top}$)
        \ENDFOR
        \FORALL{$r_k \in D_1$ }
            \STATE $r'_k$ = findClosest($d(r_k)$)
            \STATE $r''_k$ = findClosest($d(r'_k)$)
            \IF{$r''_k$ == $r_k$}
                \STATE $r_k$ and $r'_k$ are matched
            \ENDIF
        \ENDFOR
        \end{algorithmic}
\end{algorithm}
\textbf{}
\label{alg:ER}

\rev{
Verifying the symmetry of the relationship has the advantage of increasing the precision by reducing the False Positive Rate, without penalizing the recall. The effect of $n_{top}$ is described in Table \ref{tab:ntop}. }
In both algorithms, many elements (either RIDs or CIDs) will have no matches in the other dataset. 
If appropriate embeddings were learned for the RIDs, then this approach will produce good matches,
which is indeed what we observe in our experiments.

\begin{table*}[h]
\begin{tabular}{c|c|c|c|c|c|c|}
\hline
Name (shorthand) & \# tuples & \# columns & \# distinct values & \# matches & \# sentences & \% overlap \\ \hline
\hline
IMDB-Movielens (IM)         & 49875   & 15 & 118779 & 4115 & 2810900 & 8.79  \\ \hline
\rev{Amazon-Google (AG)}          & 4589    & 3  & 5390   & 1166 & 166316  & 6.01  \\ \hline
\rev{Walmart-Amazon (WA)}         & 24628   & 5  & 45454  & 961  & 1168033 & 3.10  \\ \hline
\rev{Itunes-Amazon (IA)}          & 62830   & 8  & 53079  & 131  & 1931816 & 5.84  \\ \hline
\rev{Fodors-Zagats (FZ)}          & 864     & 6  & 3282   & 109  & 69100   & 9.08  \\ \hline
DBLP-ACM (DA)               & 4910    & 7  & 6555   & 2223 & 191083  & 62.33 \\ \hline
DBLP-Scholar (DS)           & 66879   & 4  & 131099 & 5346 & 3299633 & 2.33  \\ \hline
\rev{BeerAdvo-RateBeer (BB)}      & 7345    & 4  & 11260  & 67   & 310083  & 10.18 \\ \hline
\rev{Million Songs Dataset (MSD)} & 1000000 & 5  & 870841 & 1292023   &   31180683&    n.a.   \\ \hline
\end{tabular}
    \caption{\rev{Dataset properties.}}
    \label{tab:dataStats}
\end{table*}

\stitle{Token Matching (TM).}
We also consider the problem of matching tokens that are conceptual synonyms of each other,
a task that is also known as \textit{string matching}~\cite{CADA18,ZhuHC17}. 
For example, one relation could encode a language as ``English'' while other could encode it as ``EN''.
Note that this is different from schema matching, where the objective is to identify attributes that represent the same information.
Instead, we are interested in finding pairs of \emph{tokens} from different relations that are related conceptually. 
Given two aligned attributes $A_i$ and $A_j$, we seek to identify if two tokens $t_k \in Dom(A_i)$ and $t_l \in Dom(A_j)$ are related.
Given the token $t_k$, we identify the set of top-n token ids that are closest to $t_k$.
We announce that the first token $t_l \in Dom(A_j)$ that occurs in the ranked list is the conceptual synonym of $t_k$.

\section{Experiments}
\label{sec:experiments}
\rev{In this section we first demonstrate that our proposed embeddings learn the major relationships inherent in structured data (Section~\ref{sec:EQ}). We then show the positive impact of our embeddings for multiple data integration tasks in supervised and unsupervised settings (Section~\ref{sec:DI}). Finally, we analyze the contributions of our design choices (Section~\ref{sec:ablation}).}


\stitle{Datasets.}
We used \rev{8 datasets from the literature~\cite{GokhaleDDNRSZ14,deeper,deepmatcher,DasCDNKDARP17}} and a dataset \rev{with a larger schema (IM) that} we created starting from open data (\url{https://www.imdb.com/interfaces/}, \url{https://grouplens.org/datasets/movielens/}). Details for the scenarios are in Table~\ref{tab:dataStats}. For \rev{the majority of the scenarios, less than 10\%} of the distinct data values are overlapping across the two datasets, \rev{MSD is a dataset with one table only}.

\stitle{Pre-trained Embeddings.}
In the following, \textit{pre-trained} word embeddings have been obtained  from \rev{{\sc fastText}~\cite{BojanowskiGJM17}. We tested also {\sc GloVe}~\cite{PenningtonSM14} and obtained comparable quality results.} We relied on state of the art methods to combine words in tuples and to obtain embeddings for words that are not in the pre-trained vocabulary~\cite{deeper,CakalMA19}. 


{\stitle{Embedding Generation Algorithms.}}
We test four algorithms for the generation of local embeddings from relational dataset. \final{All local methods make use of our tripartite graph and exploit record and column IDs in the integration tasks}.

The first method is {\sc Basic}, which creates embeddings from permutations of row tokens and sentences with samples of attribute tokens.
As the method is aware of the structure of the database, it can learn representation for tuples and attributes. We fixed the size of the sentence corpus for {\sc Basic} to contain the same number of tokens in \system's corpus.

The second method is {\sc Node2Vec}~\cite{node2vec},
a widely used algorithm for learning node representation on graphs. Given our graph as input, it learns vectors for all nodes.
We used the implementation from the paper with default parameters.

The third method is {\sc Harp}~\cite{ChenPHS17}, a state of the art algorithm that learns embeddings for graph nodes by preserving higher-order structural features.
This method represents general meta-strategies that build on top of existing neural algorithms to improve performance.
We used the implementation from the paper with default parameters.

The fourth method is the one presented in Section~\ref{sec:overview}, we refer to it as \system in the following (\url{https://gitlab.eurecom.fr/cappuzzo/embdi}). The default configuration uses our tripartite graph, walks (sentences) of size 60,
300 dimensions for the embeddings space, the
Skip-Gram model in word2vec with a window size of 3, \rev{and different tokenization strategies to convert cell values in nodes}.
\rev{We report the numbers of generated sentences for each dataset in Table~\ref{tab:dataStats}}. The number of sentences depends on the desired number of tokens in the corpus, we discuss a rule-of-thumb to obtain reasonable sizes in the ablation analysis.

By default, \system uses optimizations in data integration tasks.
However, to be fair to pre-trained embeddings, our default configuration does not exploit external information, therefore the techniques in Sections~\ref{sec:skolem}, \ref{sec:external}, and \ref{subsec:embeddingAlignment} are not used - we show their impact in the ablation study.
Experiments have been conducted on a laptop with a CPU Intel i7-8550U,  8x1.8GHz cores and 32GB RAM.

\begin{table*}[]
\begin{tabular}{c|c|c|c|c||c|c|c|c||c|c|c|c||c|c|c|c|}
\cline{2-17}
   & \multicolumn{4}{c||}{\sc{Basic}} & \multicolumn{4}{c||}{\sc{Node2Vec}} & \multicolumn{4}{c||}{\sc{Harp}} & \multicolumn{4}{c|}{\sc{\system}} \\ \cline{2-17}
   & \scriptsize{MA}    & \scriptsize{MR}   & \scriptsize{MC}   & \scriptsize{AVG}  & \scriptsize{MA}    & \scriptsize{MR}   & \scriptsize{MC}   & \scriptsize{AVG}  & \scriptsize{MA}    & \scriptsize{MR}   & \scriptsize{MC}   & \scriptsize{AVG}  & \scriptsize{MA}    & \scriptsize{MR}   & \scriptsize{MC}   & \scriptsize{AVG}    \\ \hline \hline
\rev{BB}  & \textbf{.99}  & .33 & .32 & .55 & .97   & \textbf{.66}   & .92   & \textbf{.85}   & .96  & .65  & \textbf{.95} & \textbf{.85} & .92  & .50  & .77 & .73 \\ \hline
\rev{WA}  & .19  & .27 & .12 & .19 & mem   & mem   & mem   & mem   & .16  & .32  & .13 & .20 & \textbf{.94}  & \textbf{1.00}    & \textbf{.99} & \textbf{.98} \\ \hline
\rev{AG}  & \textbf{1.00}    & \textbf{.42} & .10 & .51 & \textbf{1.00}  & .39   & \textbf{1.00}  & \textbf{.80}   & .99  & .37  &  \textbf{1.00} & .79 &  \textbf{1.00}   & .38  & \textbf{1.00}   & .79 \\ \hline
\rev{FZ}  & .08  & .30 & .00 & .13 & .84   & .88   & .62   & .78   & .80  & .86  & .89 & .85 & \textbf{.94}  & \textbf{.99}  & \textbf{.94} & \textbf{.95} \\ \hline
\rev{IA}  & .09  & .11 & .09 & .09 & mem   & mem   & mem   & mem   & .81  & .59  & .96 & .78 & \textbf{.89}  & \textbf{.85}  & \textbf{.98} & \textbf{.90} \\ \hline
DA        & .08  & .29 & .02 & .13 & \textbf{.79}   & .77   & .18   & .58   & .51  & .74  & .49 & .58 & \textbf{.79}  & \textbf{.91}  & \textbf{.66} & \textbf{.79} \\ \hline
DS        &  \textbf{1.00}   & .58 & .69 & .76 & mem   & mem   & mem   & mem   & .12  & .06  & .06 & .08 & \textbf{.90}  & \textbf{.99}  & \textbf{.99} & \textbf{.96} \\ \hline
IM        & \textbf{.99}  & .34 & .64 & \textbf{.66} & mem   & mem   & mem   & mem   & .07  & .29  & .10 & .16 & .74  & \textbf{.42}  & \textbf{.78} & .65 \\ \hline
\rev{MSD} & .31  & .37 & .51 & .39 & mem   & mem   & mem   & mem   & t.o. & t.o. & t.o. & t.o. & \textbf{.60}  & \textbf{.95}  & \textbf{.83} & \textbf{.79} \\ \hline

\end{tabular}
\caption{\rev{Quality results for local embeddings generation}.}
\label{tab:full_EQ}
\textbf{}
\end{table*}

\subsection{Evaluating Embeddings Quality} \label{sec:EQ}
\rev{We introduce three kinds of tests to measure how well embeddings 
learn the relationships inherent in the relational data}.
Each test consists of a set of tokens taken from the dataset as input, while the goal is
to identify which token does not belong to the set 
(function \textit{doesnt\_match} in Python library \textit{gensim}). For the \textit{MatchAttribute} (MA) tests, we randomly sample four values from 
an attribute and a fifth value from a different attribute \rev{at random} in the same dataset, e.g., given (Rambo III, The matrix, E.T., A star is born, \textit{\textbf{M. Douglas}}), the test is passed if M. Douglas is identified. In \textit{MatchRow} (MR), we  pick all tokens from a row and replace one of them at random with a value from a different row, \rev{also selected at random from the same dataset}, e.g., (S. Stallone, Rambo III, \textit{\textbf{1952}}, P. MacDonald). Finally, in \textit{MatchConcept} (MC), we model more subtle relationships. \rev{We manually identify two attributes $A_1$ and $A_2$ that are in a one to many relationship}. For a random token $x$ in $A_1$, we identify all tuples $T$ such that ($A_1=x$), we take three $A_2$ distinct values in $T$ and we finally add a random value $y$ \rev{(not in $T$)} from $A_2$.
The test is passed if $y$ is identified as unrelated from the other tokens, e.g., (Q. Tarantino, Pulp fiction, Kill Bill, Jackie Brown, \textit{\textbf{Titanic}}). This test observes whether the relationship between co-occurring elements (such as directors and their movies) is stronger than the relationship between elements that belong to the same attribute.
\rev{We took the union of the (aligned) datasets for each scenario and created between 1000 and 11000 tests, depending on its size in terms of rows and attributes.}

We report the quality results in Table~\ref{tab:full_EQ}, where each number represents the fraction of passed tests. With large datasets, some methods either failed the execution or have been stopped after a cut-off time of 10 hours.
While on average the local embeddings generated by \system are superior to all other methods, our solution is beaten in few cases. By increasing the percentage of row permutations in {\sc Basic}, results improve for MR but decrease for MA, without significant benefit for MC.  This shows that complex relationships are not modelled by row and attribute co-occurrence.  
\rev{{\sc Node2Vec} fails on our configuration for the larger scenarios with memory errors (mem), while {\sc Harp} has been stopped after 10 hours for MSD (t.o.). We do not report results for pre-trained embedding as they are not aware of the relationships in the dataset and perform very poorly for this task. For example, they obtain .33 on average for dataset BB (MA: .49, MR: .27, MC: .24) and 0.16 on average for dataset AG (MA: .03, MR: .22, MC: .22).
}

\vspace{1ex} \noindent \rev{\textit{Take-away}: our graph preserves the structure of the dataset and \system generates local embeddings that model column, row, and inter-tuple relationships better than other embedding generation methods.}

\begin{table}[h!]
\begin{tabular}{c|l|c|c|c|c|c|}
\cline{2-7}
& \multicolumn{6}{c|}{Unsupervised}\\ \cline{2-7}
& \textsc{\final{Base}} & \system & \textsc{Node2Vec} & \textsc{Harp} & SEEP$_P$ & SEEP$_L$ \\ \hline
\rev{BB} & \textbf{\final{1.00}} & \textbf{1.00} & \textbf{1.00} & \textbf{{1.00}} & .75 & .75 \\ \hline
\rev{WA} & \textbf{\final{1.00}} & \textbf{1.00} & mem & .60 & .60 & .80 \\ \hline
\rev{AG} & \textbf{\final{1.00}} & \textbf{1.00} & \textbf{1.00} & \textbf{1.00} & \textbf{1.00} & \textbf{1.00} \\ \hline
\rev{FZ} & \textbf{\final{1.00}} & \textbf{1.00} & \textbf{1.00} & \textbf{1.00} & \textbf{1.00} & \textbf{1.00} \\ \hline
\rev{IA} & \textbf{\final{1.00}} & \textbf{1.00} & mem & \textbf{1.00} & .50 & .75 \\ \hline
DA & \textbf{\final{1.00}} & \textbf{1.00} & mem & .50 & .75 & .81 \\ \hline
DS & \textbf{\final{1.00}} & .50 & mem & \textbf{1.00} & .60 & {.73} \\ \hline
IM & \final{.60} & \textbf{.78}  & mem & \textbf{.78}  & .68 & .75 \\ \hline
\end{tabular}

\caption{\rev{F-Measure results for Schema Matching (SM)}.}
\vspace{-3ex}
\label{tab:full_SM}
\end{table}

\subsection{Data Integration Tasks}
\label{sec:DI}
\rev{We 
test schema matching and entity resolution in every integration scenario with two datasets and report preliminary results on token matching.}
%
In the following, we measure the quality of the results w.r.t. hand crafted ground truth for each task with precision, recall, and their combination (F-measure). Execution times are reported in seconds.



\begin{table*}[h]
\begin{tabular}{c|c|c|c|c|c|c||c|c|c|c|}
\cline{2-11}
   & \multicolumn{6}{c||}{Unsupervised} & \multicolumn{2}{c|}{Supervised} & \multicolumn{2}{c|}{Task specific} \\
   \cline{2-7}
   & Pre-trained &  \multicolumn{5}{c||}{Local} &  \multicolumn{2}{c|}{(5\% labelled)} & \multicolumn{2}{c|}{(5\% labelled)} \\
   \cline{2-11}
      & \sc{fastText} & \sc{EmbDI-S} & \sc{EmbDI-F} & \sc{EmbDI-O} & \sc{Node2Vec} & \sc{Harp}                                & \sc{DeepER$_P$}        & \sc{DeepER$_L$}        & \sc{DeepER$_P$}          &  \sc{DeepER$_L$} \\
   \hline
\rev{BB} & .59      & .50    & .82   & \textbf{.86}     & \textbf{.86}    & \textbf{.86}  & 0.51           & 0.53          & 0.54             & 0.58            \\ \hline
\rev{WA} & .58      & .59    & .75   & \textbf{.81}     & mem       & .78  & 0.58           & 0.62           & 0.62             & 0.63            \\ \hline
\rev{AG} & .18      & .14    & .57   & .59     & .70    & \textbf{.71}  & 0.53           & 0.56           & 0.58             & 0.62            \\ \hline
\rev{FZ} & .99      & .98    & .99   & .99     & \textbf{1.00}    & \textbf{1.00} & \textbf{1.00}           & \textbf{1.00}           & \textbf{1.00}             & \textbf{1.00}            \\ \hline
\rev{IA} & .10      & .09     & .09    & .11     & mem       & .14     & .76           & .81           & .77             & \textbf{0.82}            \\ \hline
DA & .72      & .95    & .94   & .95     & .87    & \textbf{.97}  & .84           & .89           & .86             & .90            \\ \hline
DS & .80      & .85    & .75  & \textbf{.92}     & mem       & .81     & .80           & .87          & .82             & .91            \\ \hline
IM & .31      & .90    & .64   & .94     & mem       & \textbf{.95}  & .82           & .88           & .84             & .91            \\ \hline
\end{tabular}

    \caption{\rev{F-Measure results for Entity Resolution (ER)}.}
\label{tab:full_ER}
\vspace{-1ex}
\end{table*}

\stitle{Schema Matching.}
We test an unsupervised setting using  Algorithm~\ref{alg:SM} \final{with overlap of columns treated as bag-of-words ({\sc Base})} and with local embeddings. We also report for an existing system with both pre-trained embeddings ({\sc Seep$_P$}), as in the original paper~\cite{seepingSemantics}, and {\sc EmbDI} local embeddings ({\sc Seep$_L$}), as they are the ones with competitive performance that we could generate in all cases.

Table~\ref{tab:full_SM} reports the results w.r.t. manually defined attribute matches. All methods are unsupervised, but we distinguish two groups. In the first group, local embeddings are generated and then used with Algorithm~\ref{alg:SM} from Section~\ref{sec:embeddingsForIntegration}.
{\sc Basic} local embeddings lead to 0 attribute matches in this experiment \rev{and we do not report them in the table}.
While \system embeddings lead to the best results in most cases, for \rev{DS} {\sc Harp} {gets better results}. While we can get comparable results with optimizations (Section~\ref{sec:extensions}), \final{this shows that our graph enables other, more complex embedding schemes to get good results. {\sc Base} performs very well across most datasets and it is outperformed by local embeddings in one case.}

In the second group, we compare pre-trained and \system embeddings with an existing matching system. We have two main remarks.
First, the simple unsupervised method with \system embeddings outperforms {by at least an absolute 10\% the {\sc Seep$_P$} baseline in terms of F-measure in all scenarios.
Second, the baseline method improves by an average absolute 6\% in F-measure when it is executed with \system embeddings, showing their superior quality for SM w.r.t. pre-trained ones}.

We observe that results for {\sc Seep$_P$} depend on the quality of the original attribute labels. If we replace the original (expressive and correct) labels with synthetic ones, {\sc Seep$_P$} obtains F-measure values between {.30 and .38}. Local embeddings from \system do not depend on the presence of the attribute labels.
Finally, we tested a traditional instance-based schema matcher that does not use embeddings~\cite{MassmannRAAR11,MarnetteMPRS11}, whose results are lower than the ones obtained by \system in all scenarios.

\vspace{1ex} \noindent \rev{\textit{Take-away}: \system local embeddings are more effective than pre-trained ones for the schema matching task when tested with two different unsupervised algorithms.}

\stitle{Entity Resolution.}
For ER, we study both unsupervised and supervised settings. To enable baselines to execute this scenario, we aligned the attributes with the ground truth. \system can handle the original scenario where the schemas have not been aligned with a limited decrease in ER quality. 

As baseline for the \textit{unsupervised} case, we use Algorithm~\ref{alg:ER} with pre-trained embeddings \rev{({\sc fastText})}.
\rev{We report for our integration algorithm with \system embeddings in three variants of the way in which we tokenize the cell values in the dataset. {\sc EmbDI-S} (Simple) uses the original value as a token node in the graph (e.g., ``iPad 4th 2012''), while {\sc EmbDI-F} (Flatten) models it as single words (e.g., nodes ``iPad``, ``4th'', ``2012'' connected to the same RID and to the same CID). The first strategy is more accurate in the modeling on tokens with more than one word as each token gets its own embedding; this is more precise than the one derived from combining the embeddings of the single words.
However, a finer granularity is mandatory for heterogeneous datasets with long texts in the cell values for two reasons. First, accurate node merging is challenging with long sequences of words. Second, in different datasets the same entities can be split across attributes or grouped in one attribute. As an example, the BB datasets contain attributes ``beer name" and ``brewing company" but in one dataset oftentimes the name of the brewing company appears in the beer name ``brewing\_company\_A beer\_name\_1", while in the other dataset only beer\_name\_1 appears in the name column. As we do not assume any user-defined pre-processing of the attribute values, modeling the words individually is beneficial in these cases. The third tokenization strategy, {\sc EmbDI-O} (Overlap) is a trade off between the two that preserves as token nodes the cell values that are overlapping across the two datasets and models as single words the others.}

We also test our local embeddings in the \textit{supervised} setting with a state of the art ER system (DeepER$_L$), comparing its results to the ones obtained with pre-trained embeddings (DeepER$_P$).

\rev{Results in Table~\ref{tab:full_ER} for unsupervised settings show that {\sc EmbDI-O} embeddings obtain the best quality results in three scenarios and second to the best 
in four cases. In every case, local embeddings obtained from our graph outperform pre-trained ones.} For supervised settings, as in the SM experiments, using local embeddings instead of pre-trained ones increases 
the quality of an existing system. In this case, supervised DeepER shows an {average 5\% absolute improvement in F-measure
with 5\% of the ground truth passed as training data. The improvements decrease to 4\% with more training data (10\%).}
Similarly to SM, in the ER case local embeddings obtained with the {\sc Basic} method lead to
0 rows matched.


\begin{table*}[t]
\begin{tabular}{c|c|c|c|c|c|c|c|c|c|c|c|c|c|c|c|c|c|c|}
\hline
\multirow{2}{*}{$n_{top}$} & \multicolumn{6}{c|}{P} & \multicolumn{6}{c|}{R} & \multicolumn{6}{c|}{F} \\ \cline{2-19}
 & AG & BB & DA & \final{IA} & IM & WA & AG & BB & DA & \final{IA} & IM & WA & AG & BB & DA & \final{IA} & IM & WA \\ \hline
1 & \textbf{.803} & \textbf{.929} & \textbf{.991} & \final{\textbf{.278}} & \textbf{.973} & \textbf{.925} & .407 & .765 & .884 & \final{.039} & .862 & .634 & .540 & \textbf{.839} & .935 & \final{.068} & .914 & .752 \\ \hline
5 & .716 & .885 & .986 & \final{.132} & .963 & .853 & .494 & \textbf{.794} & \textbf{.917} & \final{.055} & .911 & .748 & .585 & .837 & \textbf{.950} & \final{.077} & \textbf{.936} & \textbf{.797} \\ \hline
10 & .715 & .885 & .986 & \final{.137} & .963 & .841 & \textbf{.496} & \textbf{.794} & \textbf{.917} & \final{\textbf{.078}} & \textbf{.912} & .757 & \textbf{.586} & .837 & \textbf{.950} & \final{\textbf{.100}} & \textbf{.936} & \textbf{.797} \\ \hline
100 & .714 & .885 & .986 & \final{.125} & .962 & .834 & \textbf{.496} & \textbf{.794} & \textbf{.917} & \final{\textbf{.078}} & \textbf{.912} & \textbf{.764} & .585 & .837 & \textbf{.950} & \final{.096} & \textbf{.936} & \textbf{.797} \\ \hline
\end{tabular}
    \caption{Effects of $n_{top}$ on ER quality.}
    \label{tab:ntop}
\end{table*}

Finally, we investigated if our task agnostic embeddings can be \emph{fine-tuned} for a specific task.
This process of pre-training followed by fine-tuning is a common workflow in NLP.
Specifically, we start with the relational embeddings learned by \system
and allow it to be fine-tuned for each individual tuple pair if it improves performance.
We achieve this by modifying the embedding lookup layer of DeepER.
By default, this layer does a ``lookup'' of a given token from the embedding dictionary.
We allow DeepER to learn an additional weight matrix $W$ such that the original \system embeddings can be tuned for ER. The final two columns of Table~\ref{tab:full_ER} shows the results.

\vspace{1ex} \noindent \rev{\textit{Take-away}: \system embeddings are more effective than pre-trained ones for entity resolution in both the unsupervised and the supervised settings.}

\stitle{Token Matching.}
Differently from the previous experiments, we do not claim an unsupervised solution for this integration task. In fact, we argue that our embeddings should be used as an additional signal to be combined with the other 
similarity measures used for this task, e.g.,
edit distance, Jaccard, TF/IDF.
We evaluated the accuracy of this approach on the IM scenario in matching of tokens across the two datasets in two (aligned) pairs of columns. \rev{We picked this dataset and these columns as it was possible to manually craft the ground truth for their matches.} Two columns in a pair have the information about the same entities, but expressed in different formats, such as ``DK'' for ``Denmark'', ``UK'' for ``Great Britain'', and so on. We used the unsupervised matching based on nearest neighbor also used for ER.

For the column expressing information about countries, pre-trained embeddings and Jaccard similarity obtain matches with 0.13 and 0.19 F-measure, respectively, while \system embeddings get 0.31. For the column about languages, the baselines obtain 0.17 and .20, while \system obtains 0.30. These results suggest that local embeddings can bring a stronger signal than pre-trained embeddings and Jaccard distance in string matching systems.

\subsection{Ablation Analysis}
\label{sec:ablation}
We now show the effect of the different parameters, design choices, and optimizations in our framework.

\stitle{Parameters.} Several parameters in \system affect the quality of the local embeddings. All the results reported above have been obtained using a single configuration, but the quality of the results for the different tasks increases significantly by tuning the parameters for the specific tasks.


The default setting uses
walks of size 60,
300 dimensions for the embeddings space, and the
Skip-Gram model in word2vec with a window size of 3.
We noticed that CBOW performs better than Skip-Gram on the ER task, while having  worse results in the EQ and SM.
For example, executing the ER task with CBOW increases F-measure by at least 2 absolute points for IM 
and DS. 
Similarly, decreasing the size of the walks to 5 for the SM task raises the F-measure for DS to 1. 
This is because embeddings from shorter walks better model the value overlap across columns. As this signal drives the matching task, a lower value increases the quality of the SM matches, but reduces the quality for EQ and ER. We also observe that an even lower value (3) decreases the results also for SM, demonstrating that a semantic characterization in also needed. 
A larger window for word2vec (5) has a negative effect on all tests and all datasets. Reducing the number of dimensions has limited, mixed effects on average, thus showing that our method is robust to this parameter.

A larger corpus leads to better results in general, but we empirically observed diminishing returns after a certain size. As a rule of thumb, we fix the total number of tokens in the corpus with the following formula: \#corpus tokens=(\#dist.values+\#rows )*1000.
The number of walks is derived by dividing the number of tokens by the walk length. 



We set $n_{top}=10$ in our ER experiments; by varying $n_{top}$ we observe the expected trade-offs between $P$ and $R$, as reported in Table \ref{tab:ntop} for \final{six datasets. Results for the FZ scenario do not change with different $n_{top}$ values and results for DS are close in values and trend to those reported for DA.}


\stitle{Optimizations.}
We tested 
optimizations of the original default configuration for \system. For replacement (Section~\ref{sec:external}), we used an external dictionary for one column in each dataset, e.g., different formats of country codes. The biggest improvement is in ER with an absolute 3\% on average, while the quality is stable for SM and 
EQ. For alignment (Section~\ref{subsec:embeddingAlignment}), we fed the optimization step with the outcome of the default model, i.e., we got candidate RIDs and CIDs from a first execution and then refined the embeddings with this information. This leads to a an absolute 2\% increase in F-measure for ER, with the higher contribution coming from the better recall.

\begin{figure}[ht]
\hspace{-2ex}
\includegraphics[scale=.5]{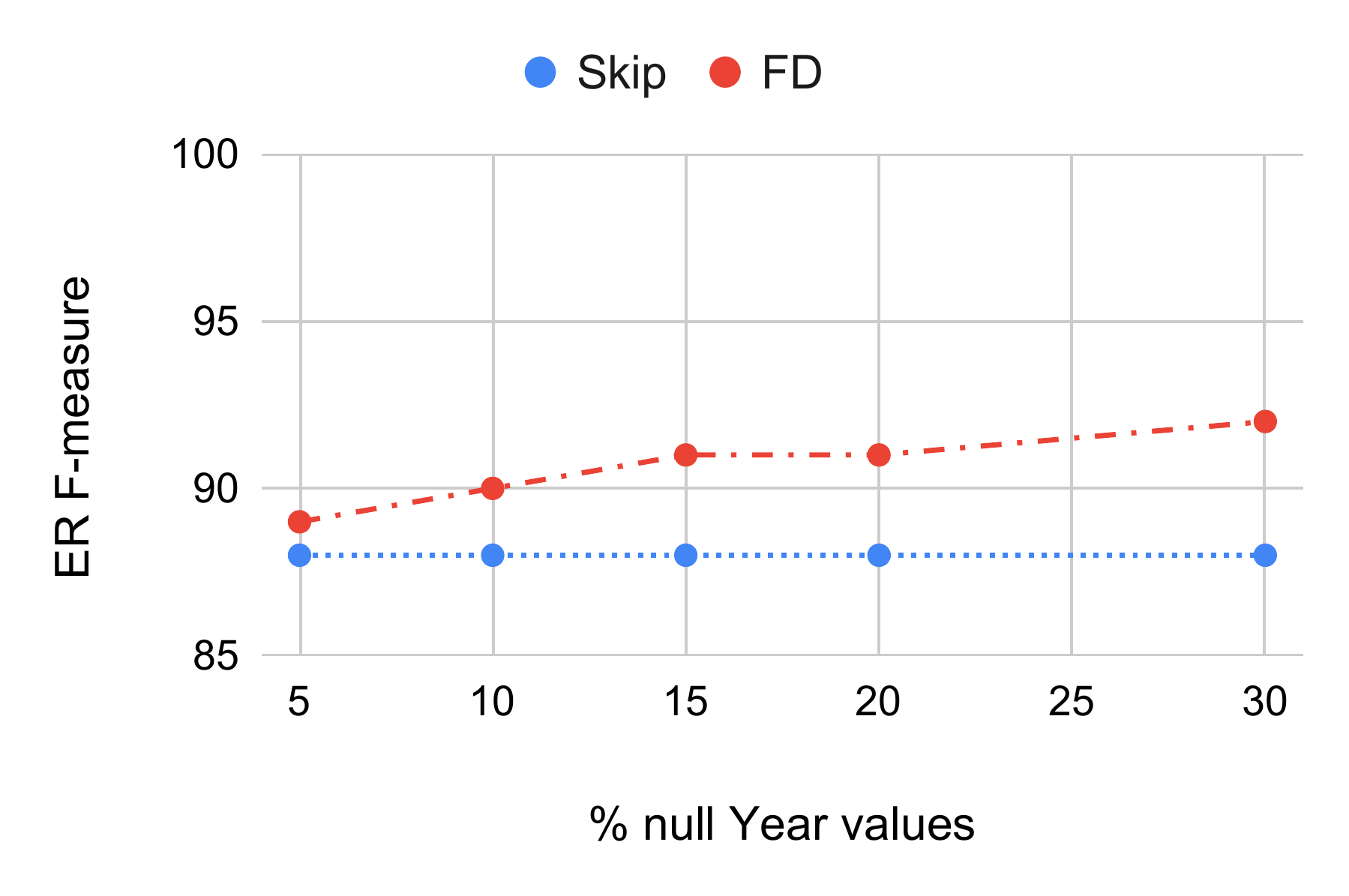}
\caption{\system ER F-measure for IM with increasing amount of missing values in the data.}
\label{fig:missing}
\end{figure}

Figure~\ref{fig:missing} shows the impact on ER of inserted missing values in the IM dataset. We defined the FD {\em Title,Director $\rightarrow$ Year} and inserted increasing amount of noise at random in the column Year. As the number of records in common across the two datasets is very low, most of the NULLs are modifying records that are only in one dataset. Surprisingly, this has a visible effects on the results in terms of F-measure. While the default Skip solution (ignore NULL values in the graph creation) stays stable until a large number of NULLs is introduced, the results improve for the optimization that enforces the FD in the graph construction. This improvement is driven by the increasing precision. In fact, there are non duplicate movies that have a large number of attribute values in common, including the year, and that are identified as duplicates by our unsupervised method (based on neighbor RIDs).
However, the FD enforces that any missing value is treated as a new value, distinct from the others, and this information moves the embedding of the RID with the NULL away from the similar tuple that is not a duplicate.


\begin{table}[ht]
\begin{tabular}{c|c|c|c|c||c|c|}
\hline
\sc{DS} & G & W & E   & W+E  & N2V & HARP    \\ \hline
\hline
\rev{BB}      & 2.47   & 66.7   & 133   & 200   & 1663   & 732   \\ \hline
\rev{WA}      & 13.4  & 329  & 1113  & 1442  & mem      & 2394  \\ \hline
\rev{AG}      & 1.19   & 34.4   & 122   & 156   & 953    & 135   \\ \hline
\rev{FZ}      & 0.3   & 12.0   & 40.7    & 52.6    & 178    & 27.0    \\ \hline
\rev{IA}      & 32.0  & 533  & 1360  & 1893  & mem      & 9122  \\ \hline
DA            & 2.08   & 43.6   & 130   & 173   & 920    & 128   \\ \hline
DS            & 33.9  & 919  & 3027  & 3947  & mem      & 21659 \\ \hline
IM            & 31.6  & 768  & 2772  & 3540  & mem      & 8001  \\ \hline
\rev{MSD}     & 146 & 6377 & 27050 & 33427 & mem      & t.o. \\ \hline
\end{tabular}

  \caption{\rev{Execution times (in seconds) for embeddings generation for \system, {\sc Node2Vec} (N2V) and {\sc Harp}.}}
\label{tab:exectime}
\end{table}

\stitle{Execution times.}
Compared to {\sc Node2Vec} and {\sc Harp}, the execution of \system is much faster and is able to compute local embeddings for all \rev{medium size} datasets in minutes on a commodity laptop.
As reported in Table~\ref{tab:exectime} for experiments with the default configuration (using word2vec and Skip-Gram), the embedding creation (E) takes on average about 80\% of the total execution time, while graph generation (G) takes less than 1\%, and sentence creation (W) the remaining 19\%. The execution times for the embeddings creation from the sentences depends drastically on the algorithm used and its configuration, e.g., CBOW is much faster than Skip-Gram.

\rev{As the graph generation is common to all methods, we compare our solution with {\sc Node2Vec} (N2V) and {\sc Harp} in terms of time to generate walks and produce embeddings (W+E). \system is faster in most cases, up to 7x in two datasets, and, \final{in contrast with {\sc Harp}}, never hits the time out (t.o.) of 10 hours. With larger datasets, {\sc Node2Vec} raised a memory error on our 32GB reference machine. \system does not suffer from this problem, even in a laptop with 16GB of main memory, we have been able to run all tests, including the ones for the biggest dataset of 1M tuples (139MB).}







\section{Next Steps}
\label{sec:conclusions}

In this paper, we proposed a novel framework -\system- for automatically learning local relation embeddings of high quality from the data.
The learned embeddings provide promising results for a number of challenging and well studied data integration tasks 
such as entity resolution and schema matching.
Our embeddings are generic to data integration and could also be tuned in a task specific manner to obtain better results. 

There are a number of intriguing research questions that we plan to tackle next. 
%
One of our key focus areas is in seamlessly combining pre-trained and local embeddings.
While blindly using pre-trained embeddings provide sub-optimal results, 
they could be intelligently combined with local embeddings provided by \system to obtain a hybrid embedding that is more effective. 
Recently, there has been increasing interest in incorporating contextual information into word embeddings and language modeling.
Approaches such as BERT~\cite{devlin2018bert} achieve state of the art results in NLP due to this.
An important open question is to formally define appropriate context for relational data integration 
so that DL models could be built for learning contextualized word embeddings. 
\subsubsection*{Acknowledgement}
This work has been partially supported by the ANR grant \textit{ANR-18-CE23-0019} and by the IMT \textit{Futur \& Ruptures} program ``AutoClean''.
\balance
\bibliographystyle{ACM-Reference-Format}
\bibliography{references}
\end{document}